# The Effect of Prosumer Duality on Power Market: Evidence from the Cournot Model

Eve Tsybina, *Student Member, IEEE*, Justin Burkett, Santiago Grijalva, *Senior Member, IEEE*

*Abstract*—Distributed energy resources behind the meter and automation systems enable traditional electricity consumers to become prosumers (producers/consumers) that can participate in peer-to-peer exchange of electricity and in retail electricity markets. Emerging prosumers can provide benefits to the system by exchanging energy and energy-related services. More importantly, they can do so in a more honest and more competitive way than the traditional producer/consumer systems. We extend the traditional Cournot model to show that the dual nature of prosumers can lead to more competitive behavior under a game theoretic scenario. We show that best response supply quantities of a prosumer are usually closer to the competitive level compared to those of a producer.

*Index Terms*—Sharing economy, smart grid, prosumers, distributed energy resources, market power, oligopoly, Cournot

## I. Introduction

ELECTRICITY MARKETS were proposed and rapidly implemented in the 1990's in various regions in the US and around the World. In 1999 and 2000, electricity markets in California and the Midwest experienced the most significant price anomalies in their history. As was later explained in [1], the main reason behind high prices during the crisis was profit seeking behavior of suppliers that obtained market power due to low price elasticity of demand. Since then, wholesale markets grew more sophisticated, and price caps and market power monitoring functions were introduced in most systems.

More recently, power markets started seeing new phenomena due to renewable energy integration and the use of consumer-side generation. Increasingly sophisticated electricity consumers have already started offering their idle assets to the market thus becoming prosumers (producers/consumers) in a "sharing economy". However, the question of broad market participation as well as possible price manipulation is still uncertain in light of these changes. In particular, the inherent properties of distributed prosumers, if carefully accounted for, may offer a promise of more competitive markets, where players bid less aggressively. The resulting markets would have higher volumes and lower prices than currently believed.

The property which affects prosumer behavior is prosumer duality. Prosumer duality is a condition in which a prosumer produces and consumes power at the same time. When a prosumer produces more than it consumes it is seen as a net producer at the point of the meter, but it is still serving a load, and hence its consumption volume is still nonzero. The prosumer still has a cost of consumption. As a result, when a prosumer attempts to maximize its producer payoff by manipulating prices, it may have to consume at those high prices, which creates a penalty on the consumer side. Similarly, if a prosumer decides to strategically adjust its behavior to maximize consumer payoff, it will have to forgo some profit it would make as a producer, which creates a similar penalty on the producer side. In some instances, like weather demand shocks, a prosumer may not know if it will clear as a net producer or a net consumer. The existence of duality provides an important balancing factor, which makes the prosumer behave less aggressively, even it is perceived as a producer or a consumer in the market.

We extend the existing theoretical base to more carefully account for prosumer duality in a market that is similar to present wholesale power markets. Prosumers are assumed to have price inelastic demand. Their supply behavior is subject to volume strategies and is modeled through Cournot oligopoly. The resulting theoretical findings and simulations allow us to see that power markets in the presence of strategic prosumers are closer to competitive level due nonzero consumption.

The rest of the paper is organized as follows. Section II provides a review of existing research on strategic prosumer behavior. Section III extends existing research to more carefully account for prosumer duality under a Cournot model. Section IV offers a numerical simulation to illustrate the theoretical findings. Section V provides an extended review of specific prosumer effects. Section VI concludes and offers an overview of further research.

## II. Literature Review

To date, significant research efforts have been conducted to investigate the mechanisms and potential effects of strategic prosumer behavior. The main concern of this study, noncooperative strategic bidding of distributed prosumers in peer-to-peer markets, has received significant attention. A review of pertinent literature is provided in Table I. We focus on objective functions of prosumers as the area which directly relates to prosumer duality.

S. Grijalva and E. Tsybina are with the School of Electrical & Computer Engineering, Georgia Institute of Technology, Atlanta, Georgia 30345 USA (e-mail: sgrijalva@ece.gatech.edu; tsybinae@gatech.edu), Justin Burkett is with the School of Economics, Georgia Institute of Technology, Atlanta, Georgia 30345 USA (e-mail: justin.burkett@gatech.edu).

TABLE I
OBJECTIVE FUNCTIONS FOR LITERATURE ON NONCOOPERATIVE STRATEGIC BIDDING OF PROSUMERS

| Objective function | Control variables | References | Limitations |
|---|---|---|---|
| min(cost of net generation + penalties from bilateral trades) | Quantity of production | [2] | Does not include duality. As a result, underestimates quantity of production and overestimates clearing price |
| min(cost of import from the grid) or max (revenue from export) | Bilateral transaction (pairs: price and quantity) | [3] | Does not include duality. Understates cost of consumption for buyers, value of production for sellers. As a result, underestimates both quantities, effect on price uncertain |
| min(price × volume of consumption) | Quantity of consumption | [4], [5] | Does not include duality. As a result, underestimates consumption quantity and clearing price |
| max(utility(consumption + distributed generation) – cost of import) or max(utility of consumption + price × minimal expected export volume) | Quantity of consumption | [6] | Does not include duality. Overestimates prices, which results in lower quantities |
| max(revenue from export – forgone utility(consumption) – cost of production) | Quantity of export | [7] | Understates the cost of consumption. As a result, mildly underestimates export quantity and mildly overestimates clearing price |
| max(volume of production – volume of consumption)×price – cost of production | Unspecified | [8] | Understates the cost of consumption. As a result, mildly underestimates export quantity and mildly overestimates market clearing price |
| max(consumer surplus – export of excess volume × price) | Price of export | [9] | This model would generate correct predictions assuming nontransitive preferences and zero production costs |
| max((price – reservation price)×quantity of supply – disutility × quantity sold^2) | Quantity of supply | [10] | Understates the value of consumption. As a result, mildly overestimates clearing price and underestimates clearing quantity |
| max(utility(production, consumption, charge and discharge, import and export from local market or the grid) + price×net export) | Production, consumption, charge, import, export quantities | [11] | The objective function as defined by equation (8) has a correct representation of utility, cost, and revenues or expenditures from transacting with the grid. But the condition from equation (5) causes the model to mildly overestimate clearing prices and underestimate clearing quantities |

Table I shows that existing objective functions do not fully account for prosumer duality. They can be grouped into three major categories depending on where they stand with respect to prosumer duality.

The first group of studies ([2], [3], [4], [5], [6]) formulates the prosumer objective function entirely without duality. Depending on the starting point (a producer or a consumer), these studies discuss strategic behavior of a prosumer whose objective function is driven purely by cost of production or cost of consumption.

The second group accounts for prosumer duality through net production and consumption volumes ([7], [8], [9]). The problem of net production and net consumption is that it allows for self-production and self-consumption at marginal cost. If a prosumer consumes at cost and imports the balance at market clearing price, its weighted average cost of consumption decreases. The result would not be the same as if it were to sell all the power it generates to the market at market clearing price and then buy the entire consumption volume at market clearing price. This becomes important when a prosumer is able to manipulate market clearing prices by leveraging its producer or consumer market power.

Two more studies ([10], [11]) account for duality and view production and consumption separately, and the only problem with them is generalizability. These two papers provide a special case of a more general formulation of prosumer objective function that will be discussed in the next subsection.

Besides the mentioned studies, there is a significant body of research addressing distributed energy sources beyond peer-to-peer markets. These include dual oligopoly or monopsony interactions with multi-agent optimization, often with price-taking consumers or producers in the face of aggregator ([12], [13], [14], [15], [16]) or optimization studies where prosumers have no export, no import, or buy and sell at a tariff ([17], [18], [19], [20], [21]). Such studies also tend to neglect duality in discussing the objective functions within distributed generation or demand response.

As has been mentioned above, most studies disregard prosumer duality or do not fully account for it, which could cause biased results. In extreme cases such as [7] such specifications of the objective function lead to contradiction in results, when a prosumer is better off rejecting its own best response function. The adjustment of models to more accurately represent prosumer duality could cause true market clearing quantities to be higher and market clearing prices to be lower than is currently believed.

III. THEORETICAL MODEL OF STRATEGIC BEHAVIOR OF A PROSUMER

A. General Model

It has been shown previously that a perfectly competitive equilibrium in an all-prosumer market is not affected by duality [22]. This paper continues the investigation of prosumer behavior to show that prosumer dual nature becomes critical under the assumption of imperfectly competitive market.

Let us suppose that prosumers can affect the market clearing price by strategically adjusting demand or supply quantities. It is now possible to see how the dual nature of a prosumer affects its strategic bidding. Since in practice producers and not consumers present the main interest for analysis and legislation, we focus on the producer effects of strategic prosumers. Let us denote by $u, c, x_b$ and $x_s$ the prosumer utility function, costs, and demand and supply quantities, respectively. The resulting market operates under the following assumptions:

1. The prosumer has a twice differentiable concave utility function such that $\frac{\partial u}{\partial x_b} > 0, \frac{\partial^2 u}{\partial x_b^2} < 0$.

2. The prosumer has a twice differentiable convex cost function such that $\frac{\partial c}{\partial x_s} > 0, \frac{\partial^2 c}{\partial x_s^2} > 0$.

3. Prosumers can estimate each other's true cost functions and are able to predict total demand.

4. Prosumers have exogenous price inelastic demand and strategic price elastic supply.

5. The market clearing price is a decreasing function of total production.

6. The transmission systems is lossless and there are not no transmission constraints that would affect the market. At clearing, the market has exactly one interior solution.

7. There are no must-run and ramping constraints. Those volumes that were not cleared in the market do not have to be produced.

8. The game is simultaneous, single stage.

9. Prosumers can affect market price unilaterally, and do not collude.

These assumptions allow to formulate best response functions resulting in a unique Cournot equilibrium, which can be found analytically as follows. We first express best response functions in the general form and then use specific best response functions to find an equilibrium.

In the general form, the total payoff of a prosumer consists of consumer payoff $u_i(x_{bi}) - p(x_{si}, x_{sj})x_{bi}$ and producer payoff $p(x_{si}, x_{sj})x_{si} - c_i(x_{si})$ and is represented by (1).

$$v_i = u_i(x_{bi}) - p(x_{si}, x_{sj})x_{bi} + p(x_{si}, x_{sj})x_{si} - c_i(x_{si}) \quad (1)$$

where
$i, j$ are the subscripts indicating $i$th and $j$th prosumer
$v_i$ is prosumer payoff
$u_{bi}$ is utility function, $c_{si}$ is cost function
$x_{bi}$ is quantity of demand of prosumer $i$, $x_{si}$ is quantity of supply of prosumer $i$
$p$ is market clearing price as a function of supply quantities of all prosumers in the market.

Maximisation of the payoff results is a first order condition for each prosumer as described in (2). Solving for $x_{si}$ results in an implicit formulation of best response supply volume (3).

$$\frac{\partial v_i}{\partial x_{si}} = -\frac{\partial p}{\partial x_{si}} x_{bi} + \frac{\partial p}{\partial x_{si}} x_{si} + p(x_{si}, x_{sj}) - \frac{\partial c_i}{\partial x_{si}} = 0 \quad (2)$$

$$x_{si} = \frac{\frac{\partial c_i}{\partial x_{si}} - p(x_{si}, x_{sj})}{\frac{\partial p}{\partial x_{si}}} + x_{bi}, \quad (3)$$

Equation (3) shows that the optimal supply that a prosumer bids to the market has two components. The first component is the traditional supply. The prosumer's own marginal cost of production $\frac{\partial c_i}{\partial x_{si}}$ in combination with price gives $p - \frac{\partial c_i}{\partial x_{si}}$, the profit markup. The effect of production cost further depends on price sensitivity to additional supply, $\frac{\partial p}{\partial x_{si}}$. These factors are the same as in a conventional Cournot equation.

The second component is not found in the absence of prosumer duality. $x_{bi}$ is the consumption volume, which adds the indirect effect originating from the dual nature of a prosumer. It stimulates prosumer $i$ to supply more to the market, in the absence of anticipation of prosumer $j$. Adding the consumption effect to the equation would drive the strategic volume of supply up. As a next step, the anticipation of $x_{sj}$ in $p(x_{si}, x_{sj})$ would cause prosumer $i$ to account for $x_{bj}$. This $x_{bj}$ would cause $x_{sj}$ to increase, which allows prosumer $i$ to withdraw part of its own supply and rely on prosumer $j$ to produce more.

### B. Detailed Model

A detailed discussion of the mentioned components requires an explicit formulation of the relationship between prices and quantities and an explicit formulation of the objective function. Let us assume that a local prosumer market operates in a single bus system illustrated in Fig. 1.

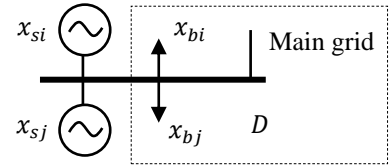

Fig. 1. The single-bus two-prosumer market.

Each prosumer contributes a strategic volume of supply to the bus, and the two prosumers are the only source of supply in the system. Further, $D$ is the total exogenous demand from the main grid faced by prosumers. It includes own demand of prosumers $x_{bi}$ and $x_{bj}$ as well as additional load to ensure that prosumer have significant market power. It should be noted that the Cournot model requires that $D > x_{si} + x_{sj}$ and in the equilibrium some demand has to be unsupplied. To prevent this this assumption from interfering with the physical nature of electricity, we model $D$ as a slack variable that can shed unsupplied load from the main grid without penalty.

Price is a decreasing function of total production, determined by the linear equation $p = D - x_{sj} - x_{si}$. Production cost is assumed to be quadratic and monotonically increasing, so that cost functions satisfy assumption 2. In the specific model, (1) takes the form of (4).

$$v_i = u_i(x_{bi}) - (D - x_{si} - x_{sj})x_{bi} + (D - x_{si} - x_{sj})x_{si} \\ - (a_{si}x_{si}^2 + b_{si}x_{si}), \quad (4)$$

where
$D$ is exogenous market demand
$a_{si}$ and $b_{si}$ are individual coefficients representing the cost of production.

The first order conditions for the market in question are given by (5).

$$x_{si}(2 + 2a_{si}) = D - x_{sj} - b_{si} + x_{bi}$$
$$x_{sj}(2 + 2a_{sj}) = D - x_{si} - b_{sj} + x_{bj} \quad (5)$$

Substitution yields a pair of equilibrium quantities expressed entirely in terms of exogenous characteristics of supply and demand, (6).

$$x_{si}^* = \frac{-2a_{sj}b_{si} - 2b_{si} + b_{sj}}{3 + 4a_{si} + 4a_{sj} + 4a_{si}a_{sj}}$$
$$+ \frac{D + 2a_{sj}D + 2a_{sj}x_{bi} + 2x_{bi} - x_{bj}}{3 + 4a_{si} + 4a_{sj} + 4a_{si}a_{sj}}$$
$$x_{sj}^* = \frac{-2a_{si}b_{sj} - 2b_{sj} + b_{si}}{3 + 4a_{si} + 4a_{sj} + 4a_{si}a_{sj}} \quad (6)$$
$$+ \frac{D + 2a_{si}D + 2a_{si}x_{bj} + 2x_{bj} - x_{bi}}{3 + 4a_{si} + 4a_{sj} + 4a_{si}a_{sj}}$$

It can be seen from the equations that the best response quantities provided in (6) can be divided into two parts.

The first fraction relates strategic supply volume to marginal costs of production for both prosumers. It is in line with predictions of the conventional Cournot model for pure producers. The own cost of production causes a prosumer to supply less, while competitor cost of production causes a prosumer to supply more.

The second fraction reflects the dependence of supply on anticipated consumption volumes. We will dedicate some additional attention to this part of (6), since it introduces prosumer-specific novelty into the equation.

In the traditional all-producer Cournot formulation, the total exogenous demand is $D_{base}$ and producers choose the supply volumes $x_{s1\,base}$ and $x_{s2\,base}$. In this case, the best response supply quantities are

$$x_{si\,base} = \frac{D_{base} - x_{sj} - b_{si}}{2 + 2a_{si}}$$
$$x_{sj\,base} = \frac{D_{base} - x_{si} - b_{sj}}{2 + 2a_{sj}} \quad (7)$$

Or, after substitution, as (8),

$$x_{si\,base}^* = \frac{-2a_{sj}b_{si} - 2b_{si} + b_{sj}}{3 + 4a_{si} + 4a_{sj} + 4a_{si}a_{sj}} + \frac{D_{base} + 2a_{sj}D_{base}}{3 + 4a_{si} + 4a_{sj} + 4a_{si}a_{sj}}$$
$$x_{sj\,base}^* = \frac{-2a_{si}b_{sj} - 2b_{sj} + b_{si}}{3 + 4a_{si} + 4a_{sj} + 4a_{si}a_{sj}} + \frac{D_{base} + 2a_{si}D_{base}}{3 + 4a_{si} + 4a_{sj} + 4a_{si}a_{sj}} \quad (8)$$

The difference found in the second fraction of (6) and (8) shows how duality affects prosumer decision-making as opposed to producer decision-making.

We compare the prosumer strategy in (6) to the baseline all-producer strategy in (8) holding the total exogenous demand constant across the markets, meaning $D = D_{base}$. A prosumer supplies more to the market whenever $D + 2a_{sj}D + 2a_{sj}x_{bi} + 2x_{bi} - x_{bj} > D_{base} + 2a_{sj}D_{base}$, which simplifies to $2a_{sj}x_{bi} + 2x_{bi} > x_{bj}$ when $D = D_{base}$. If prosumer $i$ anticipates $2a_{sj}x_{bi} + 2x_{bi} > x_{bj}$, or that it will be a relatively large consumer compared to prosumer $j$, it will produce more and manipulate market clearing price downwards. But if prosumer $i$ observes that consumption of prosumer $j$ will be much higher given its production technology, it will sacrifice its own consumer interests for the sake of making a higher profit due to strategic producer payoff. The indifference line which separates the prosumer's willingness to supply less or more is defined by (9).

$$x_{bi} = \frac{x_{bj}}{2a_{sj}+2} \quad (9)$$

where

$x_{bi}, x_{bj}$ are consumption quantities
$a_{sj}$ is the variable part of marginal cost of production for prosumer $j$.

In the area above the line, prosumer $i$ is willing to sell more than if it were a pure producer. In this half plane, prosumer $i$ anticipates that it will consume so much that the penalty from higher market prices would outweigh the additional profit of selling at a higher price. At the same time, the additional volumes provided to the market by prosumer $j$ will not stimulate prosumer $j$ to increase its quantity sufficiently to bring the equilibrium price down. As a result, prosumer $i$ finds it necessary to supply more.

It is clear that the additional volume prosumer $i$ needs to supply to the market is affected by the additional volume it anticipates from prosumer $j$. That volume, in turn, is affected by how expensive it is for prosumer $j$ to produce, determined by the slope of marginal cost $a_{sj}$. We can notice that the larger $a_{sj}$, the more area is left in the upper half plane. If $a_{sj}$ is close to 0, the line is close to 1/2, and prosumer $i$ supplies more in exactly half of the coordinate plane. If $a_{sj}$ is very large, the line approaches $x_{bj}$ axis, and prosumer $i$ supplies more on the entire coordinate plane. Fig. 2 illustrates the dependence for changing values of $a_{sj}$.

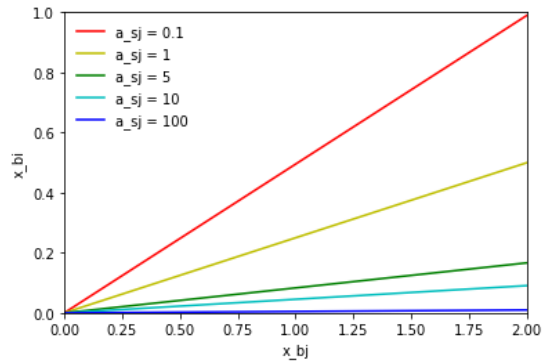

Fig. 2. The relationship between prosumer $i$'s willingness to supply and the slope of marginal costs of prosumer $j$.

For $a_{sj} = 1$ (yellow) prosumer $i$ supplies more on about three quarters of the coordinate plane. For $a_{sj} = 10$ (blue) prosumer $i$ supplies more on almost the entire plane. This finding is intuitive. If marginal costs of production increase in $x_{sj}$ very quickly, prosumer $j$ has a very high cost production technology. It cannot be relied on to reduce the market clearing price. Prosumer $i$ is likely to face a higher price in the market due to the expensive technology of prosumer $j$. To compensate for this high price, prosumer $i$ would need to add some extra production. For larger values of $a_{sj}$ it is very difficult for prosumer $i$ to supply less.

## IV. COMPARISON OF RESULTS UNDER THE PROPOSED OBJECTIVE FUNCTION AND AN EXISTING OBJECTIVE FUNCTION

### A. Two-prosumer simulation

It is possible to compare equilibrium quantities and prices resulting from both equation systems using a numerical simulation. The difference between equilibrium supply quantities and price between a baseline scenario (9) and duality scenario (6) would indicate the effect of duality on strategic behavior of prosumers. Table II provides descriptive statistics for the simulated input data.

TABLE II
DESCRIPTIVE STATISTICS OF INPUT DATA FOR COURNOT SIMULATION

|     | D    | $b_{si}$ | $a_{si}$ | $b_{sj}$ | $a_{sj}$ | $x_{bi}$ | $x_{bj}$ |
|-----|------|------|------|------|------|------|------|
| min | 5.0  | 0.0  | 0.1  | 0.0  | 0.1  | 5.0  | 0.0  |
| max | 10.0 | 5.0  | 10.0 | 5.0  | 10.0 | 10.0 | 5.0  |
| avg | 7.5  | 2.5  | 5.0  | 2.5  | 5.0  | 7.5  | 2.5  |
| N   | 1000 | 1000 | 1000 | 1000 | 1000 | 1000 | 1000 |

The inputs required for the simulation include exogenous demand $D$ and individual price inelastic demand of prosumers $x_{bi}, x_{bj}$ and coefficients to quadratic cost functions from (6) $a_{si}$ and $b_{si}$, identically and independently distributed according to a uniform distribution. Both scenarios run on the same starting conditions and the only difference between them relates to the difference in objective functions.

The outputs of the simulation include an equilibrium with market clearing quantities and prices for 1000 simulated markets for each scenario. In accordance with the previous subsection, we expect market clearing quantities and price to depend on cost functions of both prosumers. For the majority of observations, we expect quantities to be higher for duality scenario compared to the baseline scenario. Similarly, we expect market clearing price to be lower for duality scenario compared to baseline scenario.

The simulation results are provided in Table III.

TABLE III
COMPARISON OF EQUILIBRIUM SUPPLY QUANTITIES AND PRICES FOR 2-PROSUMER COURNOT SCENARIOS

|  | ALL | ABOVE THE INDIFFERENCE LINE | BELOW THE INDIFFERENCE LINE |
|---|---|---|---|
| $x^*_{si\_duality} - x^*_{si\_base}$ | 0.281 | 0.282 | -0.002 |
| $x^*_{sj\_duality} - x^*_{sj\_base}$ | 0.262 | 0.263 | -0.002 |
| $p^*_{duality} - p^*_{base}$ | -0.54 |  |  |

It can be seen that supply quantities for both prosumers tend to be higher in the duality scenario ($x^*_{si\_duality}$ is found according to (6)) and market clearing price tends to be lower in the duality scenario. When the relative contribution of expected demand of prosumer $j$ is considered (below the indifference line), supply quantities for prosumer $i$ tend to be lower, but by a very small number.

### B. Seven-prosumer simulation

In the event of more than two prosumers, the first order conditions are given by (10).

$$x_{s1}(2+2a_{s1}) + x_{s2} + \cdots + x_{sn} = D - b_{s1} + x_{b1}$$
$$x_{s2}(2+2a_{s2}) + x_{s1} + \cdots + x_{sn} = D - b_{s2} + x_{b2}$$
$$\ldots \quad (10)$$
$$x_{sn}(2+2a_{sn}) + x_{s1} + \cdots + x_{sn-1} = D - b_{sn} + x_{bn}$$

where
$i = \{1, \ldots, n\}$ are the subscripts indicating $i$th prosumer.

Alternatively, the first order conditions can be represented as a matrix equation (11).

$$\begin{pmatrix} 2+2a_{s1} & 1 & 1 \\ 1 & 2+2a_{s2} & 1 \\ 1 & 1 & 2+2a_{sn} \end{pmatrix} \bar{x}_s = D\bar{1} + \bar{x}_b - \bar{b}_s \quad (11)$$

where
$\bar{x}_s = (x_{s1} \quad \ldots \quad x_{sn})^T$ is a vector of strategic supply quantities
$\bar{x}_b = (x_{b1} \quad \ldots \quad x_{bn})^T$ is a vector of exogenous demand quantities
$\bar{b}_s = (x_{b1} \quad \ldots \quad x_{bn})^T$ is a vector of parameters representing the cost of production.

The principle behind strategic bidding of prosumers in the $n$-prosumer market stays the same. Each prosumer would strategically adjust its supply quantity $x_{si}$ to achieve the highest payoff given the anticipated behavior of other prosumers.

Let us use a 7 bus model to see how prosumer behaviour is affected by its own properties and the properties of other prosumers in the market. We start with a general case where all prosumers are initialised randomly with the values showed in Table IV. All prosumers were initialised to be equal in terms of $x_{bi}$ and $b_{si}$, thus permitting for only small variations in factors outside (7). However, $a_{si}$ was allowed to change significantly to cause a wide variety of slopes of the indifference line.

TABLE IV
DESCRIPTIVE STATISTICS OF INPUT DATA FOR COURNOT SIMULATION

|     | D    | $b_{si}$ | $a_{si}$ | $x_{bi}$ |
|-----|------|------|------|------|
| min | 20.0 | 0.1  | 1.0  | 1.0  |
| max | 30.0 | 1.0  | 10.0 | 2.0  |
| avg | 25.0 | 0.6  | 5.5  | 1.5  |
| N   | 1000 | 1000 | 1000 | 1000 |

In the presence of seven prosumers, the indifference line is represented by a 7-dimensional hyperplane making it difficult to establish if a given scenario passed above or below the indifference line for a given prosumer. But it is still possible to establish the average behaviour for each prosumer (column "ALL" in Table III). The data for average prosumer behaviour in the simulation are provided in Table V.

TABLE V
COMPARISON OF EQUILIBRIUM SUPPLY QUANTITIES AND PRICES FOR 7-PROSUMER COURNOT SCENARIOS

|  | 1 | 2 | 3 | 4 | 5 | 6 | 7 |
|---|---|---|---|---|---|---|---|
| $x^*_{si\_duality} - x^*_{si\_base}$ | 0.095 | 0.086 | 0.093 | 0.092 | 0.091 | 0.089 | 0.091 |
| $p^*_{duality} - p^*_{base}$ |  |  |  | -0.637 |  |  |  |

For every prosumer in the market, individual $x_s$ on average increased between baseline scenario and the scenario in which prosumers had their own demand $x_b$. The average market clearing price consequentially declined. This simulation certainly has its limits. It depends on the choice of starting conditions, which were chosen rather arbitrarily and in accordance with our best understanding of likely natural properties of demand and supply. But the results are in line with the predictions of the theoretical model.

## V. FURTHER INSIGHTS INTO THE PROSUMER EFFECTS

### A. Effect of production cost

Allowing for more prosumers in the market helps us better understand how the behaviour of a prosumer changes when it faces different types of competitors.

The theoretical predictions of (6) and (7) indicate that prosumer $i$ partially relies on prosumer $j$ or group of prosumers $j$ to cover their own demand $x_{bj}$. The ability of prosumer $j$ to cover own demand depends on how quickly its marginal costs increase in supply quantity $x_{sj}$. With a sufficient number of prosumers we can verify if this is indeed the case. In order to do so, we run an additional round of simulations. The scenario corresponding to (6) is tested through 8000 observations in the following way. The first 1000 observations include 7 prosumers with high cost technology. Their $b_s$ is drawn from a uniform distribution on [0.1, 1.0] and their $a_s$ is drawn from a uniform distribution on [9.0, 10.0]. For the next 1000 observations one prosumer is designed to have low cost technology. Its $a_{si}$ is drawn from a uniform distribution on [1.0, 2.0]. For the next 1000 observations one more prosumer is designed to have low cost technology. The last 1000 observations indicate a market where all 7 prosumers have low cost technology. The descriptive statistics for other simulation inputs, including $D$, $x_{bi}$, and $b_{si}$, is provided in Table VI. Both duality scenario and baseline scenario run on the same data.

TABLE VI
DESCRIPTIVE STATISTICS OF INPUT DATA FOR SIMULATION OF EFFECT OF PRODUCTION COST

|  | $D$ | $b_{si}$ | $x_{bi}$ |
|---|---|---|---|
| min | 20.0 | 0.1 | 1.0 |
| max | 30.0 | 1.0 | 2.0 |
| avg | 25.0 | 0.6 | 1.5 |
| N | 8000 | 8000 | 8000 |

The change in $x_{si}$ in the two scenarios shows how much the anticipation of competitor behaviour and resulting equilibrium prices affect the willingness of prosumer $i$ to supply more. Fig. 3 illustrates this concept of prosumer reliance on the market.

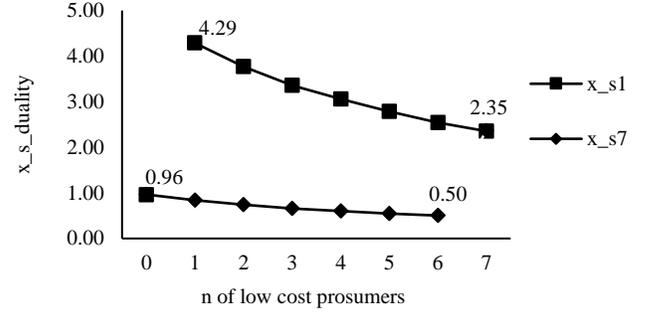

Fig. 3. The relationship between prosumer $i$'s willingness to supply versus the number of low cost prosumers $j$.

In line with the theoretical predictions, the willingness of prosumer $i$ to supply more to the market depends the ability of other prosumers to supply more. Own consumption $x_{b1}$ forces prosumer 1 to supply more to the market to prevent a high penalty on the consumer side. We start with a market in which there are no low cost prosumers. Because of high production cost, all prosumers supply relatively little. Then one prosumer, in this case prosumer 1, is simulated with low cost production technology. It realizes that it is the only low cost producer in the market and immediately increases its own production to reduce prices, $x_{s1}$=4.28. Once prosumer 2 is simulated with a low cost technology, this brings the total number of low cost prosumers to 2. Prosumer 1 is now seen to reduce its supply quantity, $x_{s1}$=3.77. Eventually, as all prosumers become low cost prosumers, $x_{s1}$ declines even further, to 2.35. This decline in production volume is robust for both high cost and low cost prosumers. In the simulated scenarios, prosumer 7 is initialized with high cost production technology. It remains a high cost prosumer throughout the simulation, even when other 6 prosumers convert to low cost. While its flexibility to supply $x_{s7}$ to the market is limited by its high cost technology, the numbers still show a decline in production. $x_{s7}$=0.96 when all prosumers are high cost and declines to $x_{s7}$=0.50 when other 6 prosumers can afford to supply more.

While it is true that a no-duality counterfactual described in (8) would also show a decrease in $x_{si}$ as the number of low cost prosumers increases, we can see that the relative drop in $x_{si}$ for a duality scenario discussed above is sharper. Fig. 4 shows that as prosumer $i$ relies more on other prosumers for supply quantity, the duality scenario gets closer to a no-duality scenario.

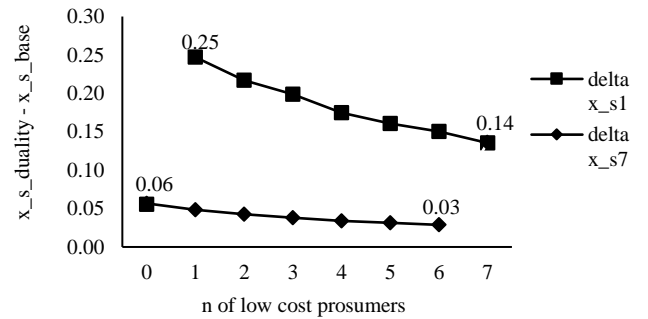

Fig. 4. The difference between prosumer $i$'s willingness to supply between two counterfactuals.

These findings represent sample averages for the generated scenarios and are in line with our discussion of Fig. 2. In markets with high $a_{sj}$ prosumer $i$ almost always supplies more, which results in higher average difference. In markets with low $a_{sj}$ which indicates low cost of production for competitors, a prosumer can expect competitors to provide large volumes of supply. It would aggressively withhold supply to strategically raise equilibrium prices, even to a point of supplying less that in the baseline scenario.

*B. Effect of demand*

The theoretical predictions of (5) stipulate that own consumption $x_{bi}$ forces prosumer $i$ to supply more to the market so that it can satisfy its own demand at reasonable prices. Larger volumes of consumption $x_{bi}$ are expected to cause higher production $x_{si}$. The same principle applies to $x_{sj}$. A larger market allows to see how this effect evolves when prosumer $i$ interacts with multiple prosumers with varying $x_{bj}$.

Similar to the previous subsection, we run 8000 additional simulations with a varying number of prosumers with high $x_b$. We start with a market in which all 7 prosumers have low $x_b$, drawn from a uniform distribution on [0.1, 1.0]. Then we design one, two and eventually seven prosumers to have a high value of $x_b$ drawn from a uniform distribution on [1.5, 2.5]. The descriptive statistics for other simulation inputs, including $D$, $a_{si}$, and $b_{si}$, is provided in Table VII.

TABLE VII
DESCRIPTIVE STATISTICS OF INPUT DATA FOR SIMULATION OF EFFECT OF DEMAND

|     | D    | $a_{si}$ | $b_{si}$ |
|-----|------|----------|----------|
| min | 20.0 | 1.0      | 0.1      |
| max | 30.0 | 2.0      | 1.0      |
| avg | 25.0 | 1.5      | 0.6      |
| N   | 8000 | 8000     | 8000     |

If prosumer $i$ expects prosumer $j$ to increase their supply $x_{sj}$ in response to larger demand $x_{bj}$, we can expect supply quantity $x_{si}$ to decline as more prosumers are simulated with high $x_{bj}$. This proves true for the simulated market. Fig. 5 shows the relationship between supply volume of prosumer $i$ and the number of prosumers with high $x_{bj}$.

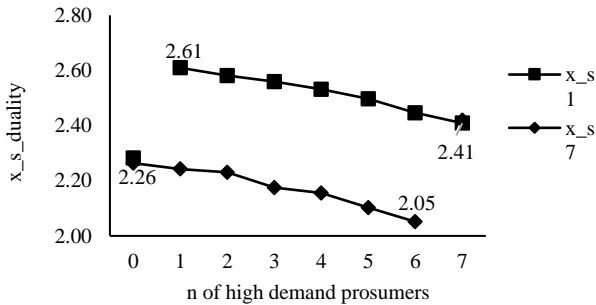

Fig. 5. The relationship between prosumer $i$'s willingness to supply and the number of prosumers $j$ with high demand.

In a market where all prosumers have low own consumption volumes, all prosumers are incentivised to supply approximately the same volume. Then one prosumer, in this instance prosumer 1, is initialised with a high $x_{b1}$. It realises that no other prosumer has a high $x_b$ and does not anticipate them to supply more. As a result, it relies on itself to manipulate price downwards through additional supply and its supply $x_{s1}$ stands at 2.61. When prosumer 2 is initialised with high $x_{b2}$, prosumer 1 expects prosumer 2 to increase $x_{s2}$. It lowers $x_{s1}$ from 2.61 initially to 2.58. As eventually all prosumers in the market have a high $x_b$, the strategic supply of prosumer 1 declines to 2.41. While prosumer 1 remains a high demand prosumer throughout the simulation, a similar behaviour pattern is observed in a low demand prosumer. Prosumer 7 is simulated as a low demand prosumer as other prosumers convert to high demand. Nevertheless, its behaviour follows the same principle as that of prosumer 1. It is first forced to supply more and $x_{s7}$=2.26, but then relies more and more on other prosumers to supply extra volume to the market. Eventually, as other 6 prosumers supply more, its strategic supply reaches 2.05.

It is also noteworthy that the effect of demand has lower contribution to changes in strategic supply volumes compared to the effect of production costs. Despite the fact that all prosumers in the simulation were initialised with relatively low values of $a_{si}$ and therefore high flexibility to change supply volumes, the change in their strategic supply was very moderate compared to that found in the previous subsection.

## VI. CONCLUSIONS

This paper describes the strategic behavior of interacting prosumers by using a Cournot model. It shows that the strategic supply quantity is affected by both production-related factors and consumption-related factors. The extent to which prosumers supply additional volume to the market depends on their own volume of consumption and their expectation of other prosumers' behavior.

These theoretical findings were tested on a narrower model with quadratic production costs. The numerical simulation suggests that indeed prosumers tend to supply more to the market if their own consumption is high or if production technology of their competitors is high cost. Further, when prosumers relied on competitors to supply more and themselves supplied less, they would withhold relatively small volume of supply. This result presents interest for additional research.

The discussion of this relationship between marginal cost, strategic supply, and resulting direct and indirect effects, reveals some shortcomings of Cournot model. Specifically, exogenous demand volumes are a special case of price inelastic demand. It may be helpful to see how prosumer strategies unfold in the presence of price elastic supply and demand. Further, it has been mentioned that market rules such as the existence of net prosumers can affect strategic behavior. Other effects can be produced by the physical nature of power markets, such as losses or transmission constraints. An extension model is needed to discuss the implications of strategic supply and demand in more detail.

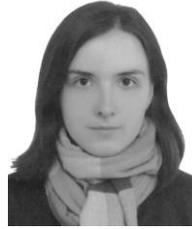

**Eve Tsybina** received Sp. degree in electrical engineering from National Research University – Moscow Power Engineering Institute, Moscow, Russia, and M.G.E. degree in International Business and Project Development from ESCP Europe, Paris, France, in 2012-2017. She is a PhD student in Georgia Institute of Technology, Atlanta.

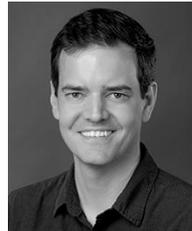

**Justin Burkett** is an Associate Professor of Economics at Georgia Tech. His research is in the area of microeconomic theory called market design, which studies the design of real-world market institutions. He received his PhD in Economics from the University of Maryland, College Park.

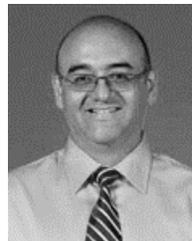

**Santiago Grijalva** is the Southern Company Distinguished Professor of Electrical and Computer Engineering and Director of the Advanced Computational Electricity Systems (ACES) Laboratory at Georgia Tech. His research interest is on decentralized power system control, cyber-physical security, and economics. From 2002 to 2009 he was with PowerWorld Corporation. From 2013 to 2014 he was with the National Renewable Energy Laboratory (NREL) as founding Director of the Power System Engineering Center (PSEC). Dr. Grijalva was a Member of the NIST Federal Smart Grid Advisory Committee. His graduate degrees in ECE are from the University of Illinois at Urbana-Champaign.